\documentclass[12pt]{article}
\newcommand{\be}{\begin{equation}}
\newcommand{\ee}{\end{equation}}
\newcommand{\ba}{\begin{eqnarray}}
\newcommand{\ea}{\end{eqnarray}}
\newcommand{\no}{\noindent}
\newcommand{\n}{\label}

\begin{document}


\title{On big rip singularities}

\author{ \large L.P. Chimento $\!^a$ and Ruth Lazkoz $\!^b$\\
{$^a$ \it  \small Dpto. de F\'\i sica, Facultad de Ciencias Exactas y Naturales,}\\
{ \it  \small Universidad de Buenos Aires, Ciudad Universitaria}\\
{ \it  \small Pabell\'on I, 1428 Buenos Aires, Argentina}\\
{\tt\small chimento@df.uba.ar}\\
{$^b$\it  \small Fisika Teorikoa eta Zientziaren Historia Saila,}\\
{\it \small Zientzia eta Teknologiaren Fakultatea, Euskal Herriko Unibertsitatea, }\\
{\it \small   644 Posta Kutxatila, 48080 Bilbao, Spain}\\
{\tt\small wtplasar@lg.ehu.es}\\}
\date{}
\maketitle
\begin{abstract}

In this comment we discuss big rip singularities occurring in typical phantom
models by violation of the weak energy condition. After that, we compare them with
future late-time singularities arising in models where the scale factor ends
in a constant value and there is no violation of the strong energy condition.
In phantom models the equation of state is well defined along the whole evolution, even
at the big rip. However, both the pressure and the energy density of the
phantom field diverge. In contrast, in the second kind of model the equation
of state is not defined at the big rip because the pressure bursts at a finite
value of the energy density.

\end{abstract}

\section{Introduction}

Investigations on theories with matter fields that violate the
weak energy condition were triggered by the influential paper of Caldwell
\cite{caldwell}, in which he showed that dark energy of that sort would fit
very well the supernova-derived luminosity. These models were dubbed phantom
cosmologies, and although phantom cosmologies have
been investigated from different perspectives, here we will only be concerned with those issues
related with analytical properties of the models \cite{earlier}. Interestingly, it has been shown that in general relativity and some of its
generalizations phantom matter might make the universe end up in a kind of
singularity \cite{bigrip} characterized by  divergences in the scale factor
$a$, the Hubble factor $H$ and its  time-derivative $\dot H$. In other words, the scale factor
expands so quickly that the scalar curvature $R$ bursts in the limit
$a\to\infty$, which is reached in a finite amount of proper time \cite{bigrip,ruth}.
These singularity is commonly called the big rip.

In a recent paper \cite{barrow}, the author suggested that the violation of
the weak energy condition would not be necessary for producing a singularity
in an expanding universe at a finite late-time. In particular, he showed that
 a singularity of that sort can be constructed in such a way that the strong weak
energy condition is preserved. Although this singularity is different than the
usual big rip, both singularities share the same physical attribute of
producing a divergent scalar curvature. This is an interesting point because
if we enlarge the definition of a big rip by saying it occurs when the scalar
curvature diverges, then we will have enriched our understanding about the meaning
of a singularity reached at finite time.  Hence, it is
interesting to clarify the nature of both types of singularities,
and we devote this comment to that purpose.

\section{Future late-time singularities}

Let us discuss the two approaches to obtaining big rip singularities. An
example of the first one can be obtained
when phantom
cosmologies are generated by a scalar field with negative kinetic term. There, the
energy density and pressure of the field are
\begin{eqnarray}
\rho_{ph}=-\frac{1}{2}\dot\phi^2+V(\phi),\\
p_{ph}=-\frac{1}{2}\dot
\phi^2-V(\phi),\end{eqnarray} while the Einstein-Klein-Gordon equations read

\be
\n{0}
3H^2=-\frac{1}{2}\dot\phi^2+V(\phi),
\ee

\be
\n{kg}
\ddot\phi+3H\dot\phi-\frac{dV}{d\phi}=0.
\ee

\no For the exponential potential

\begin{equation}
V=\frac{2(6+A^2)}{A^4}\,e^{-A\phi},
\end{equation}
with
\be
\n{fi}
\phi=\frac{2}{A}\ln{t},
\ee
the power-law evolutions
\begin{equation}
a=(-t)^{-2/A^2}\label{a_phantom},
\end{equation}

\no are solutions of the Eqs. (\ref{0})-(\ref{kg}). They represent a universe
without initial singularity beginning in the remote past with a null scale
factor. Thereafter, the scale factor grows monotonically till a future big rip
is reached at $t=0$. In our models, $\rho>0$ but the weak energy condition is
violated because $\rho+p=-\dot\phi^2<0$. So, if we associate a perfect fluid
with the phantom field, then the barotropic index for the power-law solutions
(\ref{a_phantom})

\be
\n{ga}
\gamma_{ph}=-\frac{\dot\phi^2}{\rho_{ph}}=-\frac{A^2}{3},
\ee

\no becomes negative while both the energy density $\rho_{ph}$ and the
pressure $p_{ph}$ of the field
\begin{eqnarray}
\rho_{ph}=\frac{12}{A^4 t^2}\\
p_{ph}=-\frac{4(3+A^2)}{A^4 t^2},
\end{eqnarray}
burst at $t=0$.
Note as well that the pressure
is negative, as it must, because at the big rip super-acceleration occurs.
This kind of solutions were also found by solving the semiclassical Einstein
equations for conformally invariant free fields and a conformally coupled
massive scalar field in spatially flat Friedmann-Robertson-Walker (FRW)
spacetimes with no classical radiation or matter (see Ref. \cite{an}).

In Ref. \cite{barrow} it was shown that a future finite-time singularity can
arise in as FRW expanding universe even when the strong-energy condition $\rho
+3p>0$ holds \footnote{ Recalling that for super-acceleration $p<0$ is
required, one may see that a consequence of the preservation of the
strong-energy condition is the simultaneous fulfillment of the weak-energy
condition.}. This can be done, for instance, inserting a selected scale factor with a future
finite-time singularity in the the Einstein equations for a flat FRW space

\be
\n{00}
3H^2=\rho,
\ee

\be
\n{11}
6\frac{\ddot a }{a}=-(\rho+3p),
\ee

\no we can find the energy density and the pressure satisfying the strong
energy condition. In Ref. \cite{barrow} it was chosen a model given by

\begin{equation}
a(t)=\left(\frac{t}{t_{s}}\right)^{q}\left( a_{s}-1\right) +1-\left(1-\frac{t}{t_{s}}\right)^{n},
\label{a}
\end{equation}

\no where the scale factor evolves in the interval $0<t<t_{s}$ with
$a_{s}\equiv a(t_{s})$, $1<n<2$ and $0<q\le 1$. This model represents a
universe beginning at a singularity, where the energy density $\rho$ and the
pressure $p$ are divergent while the scale factor behaves as $a(t)\approx
(a_s-1)(t/t_s)^q$, with $a_s>1$. It ends in a big rip at $t=t_s$ where the
expansion rate $H_{s}$ and energy density $\rho _{s}$ are finite positive
quantities but the pressure $p_{s}$ is a positive divergent quantity. This
type of big rip singularity is different than that appearing in phantom
cosmologies. In fact, the final behavior of the scale factor in this model is
driven by the peculiar asymptotic form of the  equation of state near $t_s$.
To see that, it will be interesting to find $p=p(\rho)$, in the limit $t\to
t_s$, using that the scale factor is a given function of the cosmological
time. To begin with, let us expand the scale factor (\ref{a}) at late times in
powers of $t_s-t$. The first correction to the constant $a_s$ is given by

\be
\n{a1}
a\approx a_s+\frac{q(1-a_s)}{t_s}(t_s-t).
\ee

\no Inserting this approximate solution in the Einstein equations (\ref{00})
and (\ref{11}) we obtain the approximate energy density, expressed in powers
of $a_s-a$, and the corresponding equation of state

\be
\n{r1}
\rho\approx\frac{3q(a_s-1)}{a_s^2t_s^2}\left[q(a_s-1)+
2n\left[\frac{a_s-a}{q(a_s-1)}\right]^{n-1}\right],
\ee

\be
\n{p1}
p\approx \frac{6n(n-1)}{a_st_s^2}\left[\frac{t_s^2}{2nq(a_s-1)}
\left[\frac{a_s^2}{3}\rho-\frac{q^2(a_s-1)^2}{t_s^2}
\right]\right]^{(n-2)/(n-1)}.
\ee

\no Hence, in the limit $a\to a_s$, we have a finite energy density

\be
\n{rol}
\rho(t)\to\rho_s=\frac{3q^2(a_s-1)^2}{a_s^2t_s^2},
\ee
and a pressure singularity
\be
\n{pel}
p(t)\to \infty,
\ee
which is of a logarithmic or a pole-like singularity depending on whether 
$(n-2)/(n-1)$ is respectively a real  or an integer number.
Equivalently, (\ref{r1}) and (\ref{p1}) correspond to a finite expansion rate $H_s=(\rho_s/3)^{1/2}$ and a
divergent $\dot H\to -\infty$, which means, an infinite acceleration $\ddot
a\to -\infty$.

Let us interpret now the peculiar fluid given by (\ref{r1})-(\ref{p1}) as a perfect fluid with equation of
state $p=(\gamma-1)\rho$. In the asymptotic limit $t\to t_s$ the 
barotropic index becomes

\be
\n{gap}
\gamma\approx
\frac{2n(n-1)a_s}{3q^2(a_s-1)^2}\left[\frac{a_s-a}{q(a_s-1)}\right]^{n-2}.
\ee

\no The exotic fluids that lead to future late-time singularities represented
by the barotropic indexes (\ref{ga}) and (\ref{gap}) do not share the usual
properties of the physical fluids i.e., $1\le\gamma\le 2$. In fact, the former
is negative $\gamma_{ph}=-A^2/3$, and the latter diverges asymptotically in the
limit $t\to t_s$.

Nevertheless, it must be stressed that the presence of a logarithmic or a pole-like singularity in the equation of state
is not a sufficient condition for a big rip singularity to occur. This can be understood
by considering a particular example of the the van der Waals fluid \cite{waals}. Its equation of state  is
\begin{equation}
p=\frac{8w\rho}{3-\rho}-3\rho^2,
\end{equation}
with $w$ a constant. Near some finite value $\rho_0$ such that $\vert \rho_0-\rho_s\vert $ is sufficiently small we  have $\vert p\vert \gg\rho$ so that we may set
\begin{equation}
p\approx\frac{8w\rho}{3-\rho}\label{vdw}
\end{equation}
and 
\begin{equation}
2\dot H\approx -p \label{ap11}\,;
\end{equation}
that is, we have a pressure dominated regime. Now, let us assume that the model enter this regime when   $\rho$ takes a  value $\rho_0$ slightly smaller (larger) than $\rho_s$. The singularity will be reached when $3H^2=\rho_s$, 
and this will only be possible if $H$ grows (decreases) with time during this pressure dominated epoch till $\rho$ reaches the value $\rho_s$. Clearly, equation (\ref{ap11}) tells us that $\rho$ will solely transit between $\rho_0$ and  $\rho_s$ if $w<0$ is negative (positive), and, in particular, for all phantom models ($w<-1$). We may view
the situation as if we had two branches of solutions, depending on the sign of $w$, and the reachability of the big rip depends on which branch is being taken.

The latter example  gives us hints for  drawing more general conclusions. Let us consider
a fluid with an equation of state of the form
\begin{equation}p\approx \frac{f(\rho)}{\vert\rho-\rho_s\vert^{\alpha}}\label{es_gen}\end{equation}
where $\alpha>0$ is a real or integer number and such that $f$ is an arbitrary function 
with definite sign and which remains finite as long as $\rho$ does too. 
The reason why we choose  expression (\ref{es_gen}) is that not only it includes 
(\ref{vdw}) in the $\rho\to\rho_s$ limit, but also (\ref{p1}).
Following an argumentation identical to the previous one, we conclude that the big rip singularity
will only be reached if $\rm{sign{(f(\rho))}}=\rm{sign{(\rho_0-\rho_s)}}$, where as before $\rho_0$ stands for the value of $\rho$ when the pressure dominated regime begins.

\section{Conclusions}

Summarizing, we believe it is worth widening the definition of big rip
singularities by considering cases in which the divergence only appears in the
pressure as those models which preserve the strong or weak energy conditions.
Nevertheless, the difference between this new way of viewing big rip
singularities and the customary one must always be kept in mind.

We have also shown that in the  examples discussed the pressure may diverge
due to a logarithmic or a pole-like singularity in the asymptotic equation of
state. Along this line we have also shown that the presence of such
singularities in the equation of state does not ensure {\it per se} that the
singularity is reachable, and we have discussed which condition must be met
for that to occur.

\section*{Acknowledgments}

L.P.C. is partially funded by the University of Buenos Aires  under
project X223, and the Consejo Nacional de Investigaciones Cient\'{\i}ficas y
T\'ecnicas.  R.L. is supported by  the University of the Basque Country through research grant 
UPV00172.310-14456/2002, by the Spanish Ministry of Science and Technology through research grant  BFM2001-0988, and  by the Basque Government through fellowship BFI01.412. 

\begin{thebibliography}{99}

\bibitem{caldwell}R.R.~Caldwell, 
Phys. Lett. B {\bf 545},  23 (2002).

\bibitem{earlier} 
S. Nojiri and S. Odintsov, Phys. Lett. B {\bf 571},  1 (2003);
D.J. Liu and X.Z. Li, Phys. Rev. D {\bf 68},  067301 (2003);  
 S. Nojiri and S.D.~Odintsov, Phys. Lett. B {\bf 562}, 147 (2003); P.~Singh, M.~Sami, and N.~Dadhich, Phys. Rev. D {\bf 68}, 023522 (2003);
J.G. Hao and X.Z. Li, Phys. Rev. D {\bf 68}, 083514 (2003); M.P.~D\c{a}browski, 
T.~Stachowiak, and M.~Szyd\-{\l }owski,  Phys. Rev. D {\bf 68}, 103519 (2003);
P.F. Gonz\'alez-D\'\i az,  Phys. Lett. B {\bf 586}, 1 (2004);
 E. Elizalde and J. Quiroga Hurtado, Mod. Phys. Lett. A {\bf 19}, 29 (2004); A.~Feinstein and S.~Jhingan,   Mod. Phys. Lett A {\bf 19}, 457 (2004);  
X.Z. Li and  J.G. Hao,
 hep-th/0303093; H. Stefancic, astro-ph/0310904;V.B. Johri, astro-ph/0311293; I. Brevik, S. Nojiri, S.D. Odintsov, and L. Vanzo,  hep-th/0401073; G. Calcagni hep-ph/0402126; S. Nojiri and S.D.~Odintsov,  astro-ph/0403622; J.G. Hao and X.Z. Li, astro-ph/0404154.

 
 \bibitem{bigrip}R.R.~Caldwell, M.~ Kamionkowski, N.N.~Weinberg, 
Phys. Rev. Lett. {\bf 91}, 071301 (2003);   P.F. Gonz\'alez-D\'\i az, Phys. Rev. D {\bf 68}, 021303 (2003);   V. Sahni and Y. Shtanov, J. Cosm. Astro. Phys. {\bf 0311}, 014 (2003);J.G. Hao and X.Z. Li, 
astro-ph/0309746;
A.  Yurov,  astro-ph/0305019; M. Sami and A. Toporensky,  gr-qc/0312009;M. Bouhmadi-L\'opez and J.A. Jim\'enez Madrid, astro-ph/0404540.
\bibitem{ruth}
L.P. Chimento and R. Lazkoz,
Phys. Rev. Lett. {\bf 91} 211301, (2003).

\bibitem{barrow}
J.D. Barrow, Class. Quan. Grav. {\bf 21}, L79 (2004).

\bibitem{an}
P.R. Anderson
Phys.Rev. D {\bf 33}, 1567 (1986).


 
\bibitem{waals}G.M. Kremer, Phys. Rev. D {\bf 68}, 123507 (2003); gr-qc/0401060.

\end{thebibliography}
\end{document}